\documentclass[12pt,a4paper]{article}
\usepackage[utf8]{inputenc}
\usepackage{amsmath, esint}
\usepackage{amsfonts}
\usepackage{amssymb}
\usepackage{graphicx}
\usepackage{yfonts}
\usepackage{mathrsfs}
\usepackage{hyperref}
\usepackage{tcolorbox} %to draw box around text
\usepackage[normalem]{ulem}
\usepackage{indentfirst}
\usepackage{listings}
\usepackage{caption}
\usepackage{booktabs}
\usepackage{caption}
\usepackage{bm}
\usepackage[left=1.5cm,right=1.5cm,top=2cm,bottom=2cm]{geometry}

\author{Ademir Xavier Jr\footnote{E-mail:xavnet2@gmail.com}.\\Brazilian Space Agency, \\Brasília, DF - Brazil}
\title{\textbf{The mass-energy relation and\\
the Doppler shift of a relativistic light source}\footnote{Pre-print version. Submitted to Physics Education, February 3 2020.}}
\date{}
\begin{document}
\maketitle

\abstract{This work considers the cospectral and arbitrary light emission of a moving source. The observed wavelengths of the emitted photons are described in term of kinematic and dynamical Doppler shifts in which the mass-energy relation plays a fundamental role. The presentation is an alternative way of emphasizing the importance of the concept of proper mass as a conserved quantity and the implications of the mass-energy relation when a source emits radiation. The physical contexts in which the source changes velocity after emission are discussed and a set of additional problems is presented.}
%\tableofcontents
\section{Introduction}
Relativity has changed the way we understand the dynamics of bodies interacting via electromagnetic radiation. In fact, the development of relativity can be seen as an attempt to unify electromagnetism and mechanics \cite{wittaker}. Since mechanics provided a wide range of applications in the two centuries that followed Newton's work and therefore was seen as a solid theoretical framework, relativity and its new world view were deep revolutionary steps after their first predictions were confirmed.  Such revolution represented the incorporation of electromagnetic laws into our understanding of the mechanical world. 

The historical context of relativity coincides with the downfall of the Ether hypothesis \cite{wittaker} as an all pervading medium responsible for the propagation of light much in the same way as the air is the medium in which sound waves propagate.  The existence of a frequency shift in the light emitted by a moving source was seen both as an evidence of this medium \cite{wittaker} as of the mechanism of light propagation through it. However, the Doppler shift is not the only effect produced by sources in motion, light aberration was recognized as an important astronomical correction in the position of stars since the XVIIIth century \cite{stewart}, because it was soon realized the Earth moves itself in relation to the alleged Ether. Disagreements between theoretical and experimental predictions for both effects soon became unsolvable [8] by the turn of the XXth century, giving rise to the new paradigm of special relativity. 

One of the basic concepts in which difficulties of understanding often arise is the idea of mass which remains constant during the temporal evolution of a system in a non-invariant description. With the idea of mass, a conservation law is associated, the so-called mass conservation law. However, in relativity, mass is neither ordinarily constant nor can be added \cite{gabo} in the same way as in non-relativistic mechanics. The primary conserved quantities in the new paradigm are momentum and energy, which are intrinsically linked through the definition of a new fundamental entity: the invariant 4-momentum. On the other hand, special relativity can be seen naively as a dynamic of bodies described by distinct reference systems moving in relation to each other at speeds approaching $c$ (the velocity of light). Most relativistic effects (in the length of objects, their velocities and clock rates) will therefore show up only when high velocities are involved.  While relativity is understood as a relevant matter for the physicist curriculum, the high velocity limit imposes serious limitations in the practical appreciation of its effects. 

The clash between past and modern ways of understanding this world reverberates until today when students have to learn the basics of relativity after being taught many concepts of non-relativistic mechanics. There is a way however of demonstrating the impact of relativity on low velocity systems which is the aim of this work. It involves radiation and the reformed concept of mass which is presented to students only later. Apart from past and modern controversies \cite{wang, hecht} involving the definition of rest mass, relativistic mass and proper mass \cite{hecht, okun, sandin}, light has a special role to play in such demonstration which is even more paradoxical since light and its associated photons are considered massless. In particular, we will be interested in the dynamics of a light source as described from two distinct reference frames and the role played by the source mass in the Doppler shift of the emitted photons. 

To illustrate the main concepts and avoid complications arising from more complex movements, we restrict the analysis to the motion of a source emitting radiation along the line of motion (forward and backward emissions). Limiting the analysis to 1-d motion further allows to illustrate the radiation process on the space-time diagram as discussed in Section 4.1.

\section{The radiation of a moving source}

The derivation of Doppler shift relations according to the principles of relativity is presented in a variety of ways in the literature \cite{french, dich}, always with the fundamental Lorentz transformation relations as the starting point. Thus, \cite{dich} considers a constraint which would render invariant the phase of a light wave emitted by a source at rest in the reference system $S_0$ as seen from another system $S$ moving with velocity $v$ (say, along $x$-axis) in relation to $S$. If $\nu_0$ is the frequency of the source in its reference frame, the frequency measured by the observer is
\begin{equation}
    \nu = \nu_0\gamma (1+\beta\cos\theta), 
\end{equation}
with $\beta = v/c$, $\gamma = (1-\beta^2)^{-1/2}$ and $\theta$ the angle between the line of motion of the source and the position of the observer. If the source moves toward the observer ($\theta =0$), Eq. 1 reduces to
\begin{equation}\label{doppler}
    \nu = \nu_0 \sqrt{\frac{1+\beta}{1-\beta}}.
\end{equation}

In these derivations, mass plays no role. The Doppler shift is seen as a kinematic effect arising from a relative state of motion between source and observer. Light emission in fact involves a change of state by the source. Since the energy is conserved for the integral system (source+observer), a change in the source mass $\Delta m$ is expected as the ratio
\begin{equation}\label{eqenequivalent}
    \Delta m = \Delta E/c^2,
\end{equation}
with $\Delta E$ the total amount of energy of the emitted radiation. Eq. \ref{eqenequivalent} follows from the famous mass-energy equivalent relation $E=Mc^2$ which establish a correspondence between the source `rest mass' (also called `proper mass')  $M$ before the emission, and the rest energy $E$. The mass-energy equivalence is obtained \cite{fey} from an integral of motion (total energy) based on a generalization of Newton's law $ F=d p/dt$ with $F$ an external force applied to the source and $p$ the relativistic momentum $p = \gamma\beta Mc$. However, the truly conserved quantity is the 4-momentum
\begin{equation}\label{4mom}
    {\cal P}=\left(\begin{matrix}p\\ iE/c\end{matrix}\right)=\left(\begin{matrix}\gamma\beta Mc\\ i\gamma Mc\end{matrix}\right),
\end{equation}
whose squared norm ${\cal P}\cdot{\cal P}=-Mc^2$ is an invariant and proportional to the system total mass. Since $c$ is large, the correction given by Eq. \ref{eqenequivalent} is considered too small to have any relation to the Doppler shift: it is a `side effect' of the internal process of light emission. Only in the limit of small masses (particles) or high-energy photons and source velocities (hence, in the works of high energy particle physics) such effects would play a relevant role. 

\subsection{The Doppler shift during cospectral emission in the source rest frame}

\begin{figure}[!t]
\centering
\includegraphics[width=0.4\columnwidth]{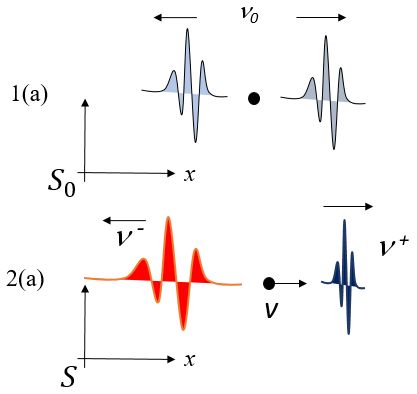}
\caption{\small Cospectral emission: 1(a) a source at rest in a laboratory frame emits two identical photons in opposite directions; 1(b) the same situation as seen from a reference system in which the source moves toward $+\hat{x}$.}
\label{fig:fig01}
\end{figure}
Restricting the description to 1-d motion for simplicity (see Fig. \ref{fig:fig01}), we consider two distinct problems:
\begin{enumerate}
    \item (a) A source with proper mass $M$ at rest in the laboratory frame emits two counter propagating photon pulses simultaneously with the \textit{same frequency} $\nu_0$. Find the final proper mass $M'$ of the source after such `cospectral emission'.
    \item (a) The same as above as described by an inertial frame moving to the left with velocity $v$.
\end{enumerate}

Using conservation of energy (C. E.), problem 1(a) is solved easily with
\begin{equation}
    M'c^2+2\epsilon_0=Mc^2,
\end{equation}
and $\epsilon_0=h\nu_0$. Because both photons have the same frequency, they carry the same momenta and the final source velocity is $v=0$. Conservation of momentum (C. M.) is implicit in such symmetrical system at rest. Therefore, the answer of problem 1(a) is 
\begin{equation}\label{masschange}
    M'=M\left(1-2z_0\right),
\end{equation}
with $z_0=h\nu_0/Mc^2$, the ratio between the one photon energy and the source rest energy. Therefore, the final proper mass of the source is reduced by the amount $2z_0$. For macroscopic bodies and low energy photons $z_0 \ll 1$ (see Section 3.1), and such mass change is `negligible'. 

In problem 2(a) the source moves initially with velocity $v\neq 0$, and the photon frequencies are $\nu_{+}$ (to the right) and $\nu_{-}$ (to the left) due to the Doppler shift. Before moving on, let us define the dimensionless quantities:
\begin{equation}\label{defin}
 \begin{split}
z^{+} & = h\nu^{+}/Mc^2, \\
z^{-} & = h\nu^{-}/Mc^2, \\
\mu & = M'/M.
\end{split}
\end{equation}

Now, C. E. demands
\begin{equation}\nonumber
    \gamma Mc^2=\gamma' M'c^2+h\nu_{+}+h\nu_{-},
\end{equation}
which, in view of Eqs. \ref{defin}, can be written as
\begin{equation}\label{eqce}
    \gamma = \mu \gamma' + z^{+}+z^{-},
\end{equation}
where again primed quantities correspond to the state after the photon emission. In the same way, C. M. requires that
\begin{equation}\nonumber
    \gamma Mv=\gamma' M'v'+\frac{h\nu_{+}}{c}-\frac{h\nu_{-}}{c},
\end{equation}
which in dimensionless variables may be rewritten as
\begin{equation}\label{eqcm}
    \gamma\beta = \mu \gamma'\beta' + z^{+}-z^{-}.
\end{equation}

 The two fundamental equations, Eqs. \ref{eqce} and \ref{eqcm}, can be be solved for $\mu$ and $\beta'$ or $z^{+}$ and $z^{-}$. However, since the context of problem (1) is given, the last option reads
 \begin{equation}\label{zs}
z^{\pm} =  \frac{1}{2}\left[\sqrt{\frac{1\pm\beta}{1\mp\beta}}-\mu\sqrt{\frac{1\pm\beta'}{1\mp\beta'}}\right]
\end{equation}
where the two equations are written in a single line using the $\pm$ symbol. Because the observer knows that $\beta'=\beta$ (in fact, the observer sees no velocity change) and, in view of Eq. \ref{masschange} or $\mu=1-2z_0$, we find from Eqs. \ref{zs}
 \begin{equation}\label{olddoppler}
 z^{\pm}  =  z_0\sqrt{\frac{1\pm\beta}{1\mp\beta}},
\end{equation}
corresponding exactly to the kinetic Doppler shift, Eq. \ref{doppler}, for each photon. According to these equations $z^{+} > z_0 > z^{-}$, because one photon is moving toward the observer while the other one is moving away from him. Thus the mass-energy relation is of fundamental importance in the origin of the kinetic Doppler effect. The mass change of Eq. \ref{masschange} is related directly to a fundamental parameter of the emitted radiation.

\subsection{The dynamical Doppler shift}

\begin{figure}[!h]
\centering
\includegraphics[width=0.4\columnwidth]{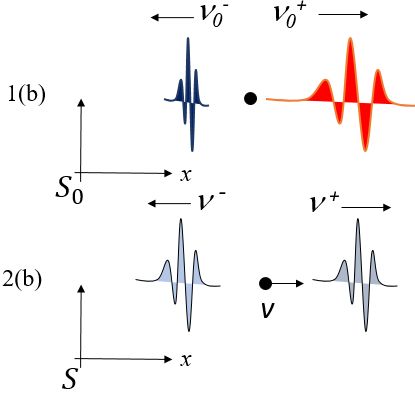}
\caption{\small Arbitrary emission: 1(b) a source at rest in a laboratory frame emits distinct photons in opposite directions; 1(b) the same situation as seen from a reference system in which the source moves toward $+\hat{x}$.}
\label{fig:fig02}
\end{figure}

In principle, the context of problem 2(a) is completely general. An observer would have no way to know that the two emitted photons have the same frequency in the source reference system - if he sees a moving source. The only thing the observer could do is to measure the photon frequencies and the final source velocity $\beta'$. If $\beta'=\beta$, the source proper frequency could be inferred by the observer from $z^{+}$ and $z^{-}$ as $z_0=\sqrt{z^{+}z^{-}}$. Hence, a new set of problems can be enunciated (Fig. \ref{fig:fig02}):
\begin{enumerate}
    \item (b) A source with proper mass $M$ at rest in the laboratory frame emits two counter propagating photon pulses simultaneously with the \textit{distinct frequencies} $\nu_0^{+}$ and $\nu_0^{-}$ to the right and to the left, respectively. Find the final proper mass $M'$ of the source after the emission.
    \item (b) The same as above as described by an inertial frame moving to the left with velocity $v$, given that the observed photon frequencies are $\nu^{+}$ and $\nu^{-}$.
\end{enumerate}
The transition from problems (a) to (b) corresponds to a dynamical change of state which distinguishes itself in principle from the purely kinematic description. The photon frequencies are a function of an internal process of radiation generation which becomes accessible externally by measuring $\nu^{\pm}$. Although no force actuates on the source, there is a velocity change, $v'\neq v$, that is, the source experiences a \textit{recoil}. If $v=0$, it may be set in motion after the emission. A solution for $\mu$  and $\beta'$ from Eqs. \ref{eqce} and \ref{eqcm} in terms of $z^{\pm}$ and $\beta$ are  obtained easily by defining
 \begin{equation} \label{defs2}
 \begin{split}
\alpha &= \sqrt{\frac{1+\beta}{1-\beta}},\\
\alpha' &= \sqrt{\frac{1+\beta'}{1-\beta'}}.
\end{split}
\end{equation}
 in terms of which, Eq. \ref{eqce} and Eq. \ref{eqcm} are written as
  \begin{equation}
 \begin{split}\label{eqscmce}
\alpha &= \mu\alpha'+2z^{+},\\
\alpha' &= \alpha(\mu+2z^{-}\alpha)'.
\end{split}
\end{equation}
It is straightforward to eliminate $\mu$ from Eqs. \ref{eqscmce} and find the new source mass
\begin{equation}\label{masschange1}
\begin{split}
\mu &=\\
& \sqrt{\left(1-2z^+\sqrt{\frac{1-\beta}{1+\beta}}\right)\left(1-2z^-\sqrt{\frac{1+\beta}{1-\beta}}\right)}.
\end{split}
\end{equation}
The answer to problem 1(b) (source at rest) is calculated by setting $\beta=0$ in Eq. \ref{masschange1} and noting that, in this reference frame, $z^{\pm}=z_0^{\pm}$ or
\begin{equation}
    \mu_{\beta=0} = \sqrt{(1-2z^+)(1-2z^-)}.
\end{equation}
If moreover $z_0^{+}=z_0^{-}$ as in problem 1(a), Eq. \ref{masschange} is retrieved. 

In order to study the source recoil, we should calculate $\beta'$. Eqs. \ref{defs2} can be rewritten as
\begin{equation}\nonumber
\begin{split}
       \beta & = \frac{\alpha^2-1}{\alpha^2+1},\\
       \beta' & = \frac{\alpha'^2-1}{\alpha'^2+1},
\end{split}
\end{equation}
and from Eqs. \ref{eqscmce}, an expression for  $\beta'$ is found as
\begin{equation}\label{velgen}
   \beta' = \frac{\gamma\beta-(z^{+}-z^{-})}{\gamma-(z^{+}+z^{-})},
\end{equation}
in terms of $\beta$ and the measured `Doppler shifts'. The recoil expression in the source initial reference frame is therefore
\begin{equation}\label{vel0}
   \beta'_{\beta=0} =- \frac{(z_0^{+}-z_0^{-})}{1-(z_0^{+}+z_0^{-})}.
\end{equation}
If $z_0^{+}>z_{0}^{-}$, $\beta'<0$, that is, if the right propagating photon is more energetic than the left propagating one, the source will recoil to the left as intuitively expected. The opposite happens if  $z_0^{+}<z_{0}^{-}$. This is the principle of the photon rocket \cite{mic,datta, singal}. The denominator of Eq. \ref{vel0} is positive because it implies in the inequality $Mc^2\ge h\nu_0^{+}+h\nu_0^{-}$ or the total photon emission energy is exhausted potentially by the total energy content of the source represented by its rest energy.

\subsection{Inverted Doppler shift}

An interesting fact about Eq. \ref{velgen} is that there is velocity change for $\beta\neq 0$ even though $z^{+}=z^{-}$ because the mass content of the source has changed. In relativistic terms, given the momentum conservation equation $\gamma'M'v'=\gamma M v$, if $M'$ is reduced by isotropic and cospestral emission, $v'$ has to increase in modulus. For $\beta \ll 1$, to first order in $z^++z^-$, the source velocity after emission is written explicitly as
\begin{equation}\label{velchangeatrest}
    v'\approx v\left[ 1+\frac{h(\nu^++\nu^-)}{Mc^2}\right].
\end{equation}
The situation seems paradoxical because, had the observer chosen originally a reference system comoving with the source, that is, a reference frame for which $v=0$, no velocity change would be observed! However, a little bit of analysis shows that the paradox is only apparent, and arise from the point of view of non-relativistic mechanics. For, in the relativistic case, if the observer chooses a frame in which $v=0$, the two photons would not be cospectral, and the velocity change would be compatible with the one calculated in Eq. \ref{velchangeatrest}. 

To see this exactly, using the Lorentz transformation for velocities \cite{french,fey}, the new velocity in the original source reference system (for $z^{\pm}=z$) will be given by
\begin{equation}\label{newvelinverteddoppler}
    \beta_0'=\frac{2z\beta\gamma}{1-2z\gamma} .
\end{equation}
In the source reference system, in accordance to Eqs. \ref{zs}, the photon dimensionless energies are then expressed as
\begin{equation}
    z_0^{\pm}=\frac{1}{2}\left[1-\mu\sqrt{\frac{1\pm\beta_0'}{1\mp\beta_0'}}\right],
\end{equation}
because $\beta_0=0$. Substituting Eq. \ref{newvelinverteddoppler} into the equation above and writing everything in terms of the original reference frame velocity $\beta$ we find
 \begin{equation}\label{newdoppler}
z_0^{\pm} =z\sqrt{\frac{1\mp\beta}{1\pm\beta}}\quad.
\end{equation}
Therefore, under special circumstances, a moving source can emit photons exhibiting no Doppler shift. The photons of the source, if observed in its proper frame, will show different frequencies in accordance to Eq. \ref{newdoppler} whose velocity ratios are inversely proportional to the ones of the original Doppler shift,  Eq. \ref{olddoppler}. Moreover, the source will show a small velocity change (as given by Eq. \ref{newvelinverteddoppler}) in its proper system, which is expected because the photons have different momenta. 

Given Eq. \ref{velgen}, it is possible to find the condition on radiation emission for which no velocity change is observed. Calling the special constant velocity $\bar{\beta}$ we find
\begin{equation}\label{velconst}
    \bar{\beta}=\frac{z^+-z^-}{z^++z^-}.
\end{equation}
Thus, for a co-moving frame with the source, $\bar{\beta}=0$ if $z_0^+=z_0^-$. Given a reference frame in which the source moves with an arbitrary velocity $\beta$, $z^\pm$ in this frame will be such that Eq. \ref{velconst} is obeyed and $\beta'=\bar{\beta}$. From this equation, one immediately obtains the mass ratio that keeps the velocity constant, $\bar{\mu}$, or
\begin{equation}
    \bar{\mu}=1-2\sqrt{z^+z^-}.
\end{equation}

%To illustrate the importance of pointing out the true meaning of mass in relativity, the after emission 4-momentum should be calculated
%\begin{equation}
%    {\cal P'}= \left(\begin{matrix}Mc[\gamma'\mu\beta'+(z^+-z^-)]\\
%    i Mc[\gamma'\mu+(z^++z^-)]\end{matrix}\right).
%\end{equation}
%Its squared norm becomes, in terms of the after-emission variables,
%\begin{equation}\label{squarednorm}
%    {\cal P'}\cdot{\cal P'}=-Mc^2\left(\mu+\frac{2z^+}{\alpha'}\right)\left(\mu+2z^-\alpha'\right).
%\end{equation}
%Now, Eqs. (\ref{eqscmce}) can be rewritten as
%  \begin{equation}
% \begin{split}\nonumber
%\frac{\alpha}{\alpha'} &= \mu+\frac{2z^{+}}{\alpha'},\\
%\frac{\alpha'}{\alpha} &= \mu+2z^{-}\alpha',
%\end{split}
%\end{equation}
%showing that the product of each expression in the parenthesis of Eq. (\ref{squarednorm}) is equal to one and ${\cal P}^2={\cal P'}^2$. Mass is strictly conserved in the process. 

\section{The limits of the radiation emission energy}

Since in practical cases $z^{\pm}\ll 1$ and $\beta\ll 1$, all relations obtained here suggest expansions in terms of these coefficients. An example is Eq. \ref{masschange1} which relates the mass change to $z^{\pm}$ and $\beta$. Considering that in practical cases the values of $z^{\pm}$ are very small, one can expand preferably Eq. \ref{masschange1} in powers of z (see the Appendix) and obtain a much simpler expression to manipulate. We should be careful, however, in using such new expansions because they may imply in mixing up concepts pertaining to distinct physical theories (e. g., classical versus relativistic dynamics because the expanded versions represent ``corrections'' to an ordinary non-relativistic behavior). Because $z^{\pm}$ is so small for current photonic propulsion systems, a similar expansion of Eq. \ref{velgen} leads to an approximate relation for the velocity gain (or loss). The resulting equations are easier to manipulate because they are linear in $z$.

To second order in $z^{\pm}$, including the velocity dependent terms, such simpler relations are
\begin{flalign}
    \mu &= 1-\left(z^+\sqrt{\frac{1-\beta}{1+\beta}}+z^-\sqrt{\frac{1+\beta}{1-\beta}}\right)-\frac{1}{2}\left(z^+\sqrt{\frac{1-\beta}{1+\beta}}-z^-\sqrt{\frac{1+\beta}{1-\beta}}\right)^2,\\
\beta'  &= \beta[1+\sqrt{1-\beta^2}(z^++z^-)]-\sqrt{1-\beta^2}(z^+-z^-)-\\
  & (1-\beta^2)(z^+-z^-)[z^+(1-\beta)+z^-(1+\beta)].\nonumber
\end{flalign}

The energy of the two emitted photons cannot be arbitrary. The emission is source-dependent and, as such, it is established by a constraint among $\nu^+$, $\nu^-$, $\beta$ and the source rest mass. Eqs. \ref{masschange1} and \ref{velgen} can be used to extract the energy restriction relations on the range of possible photon energies $h\nu^+$ and $h\nu^-$. To begin with, Eq. \ref{masschange1} is constrained by $0\le \mu \le 1$ or
\begin{equation}\label{const1}
\begin{split}
       &  z^+ \le \frac{\alpha}{2}, \\
       & z^- \le \frac{1}{2\alpha},\\
       & z^+\ge \frac{z^-\alpha^2}{2z^-\alpha-1}.
\end{split}
\end{equation}
\begin{figure}[!ht]
\centering
\includegraphics[width=0.4\columnwidth]{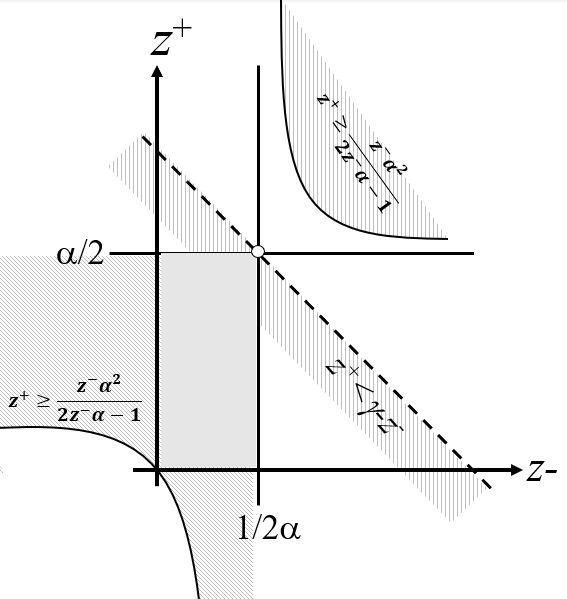}
\caption{\small Graphical representation of the feasibility region on the ($z^+,z^-$) plane for the photon emission energies as determined by Eqs. \ref{const1} and \ref{const2}.}
\label{fig:fig03}
\end{figure}
A constraint in the total energy is represented by the denominator of Eq. \ref{velgen}, or $\gamma-(z^++z^-)> 0$ or
\begin{equation}\label{const2}
    z^+ < \gamma-z^-.
\end{equation}
These energy frontiers are graphically represented in Fig. (\ref{fig:fig03}) where traced lines are the zones defined by the inequalities Eq. \ref{const1} and Eq. \ref{const2}. The third relation in Eq. \ref{const1} defines two zones above $z^+=z^-\alpha^2/(2z^-\alpha-1)$ on the $(z^+,z^-)$ plane. The interception implies simply that  $0\le z^+ \le \alpha/2$ and $0\le z^- \le 1/2\alpha$, or the squared region shown in Fig. (\ref{fig:fig03}). 

\begin{figure}[!h]
\centering
\includegraphics[width=0.4\columnwidth]{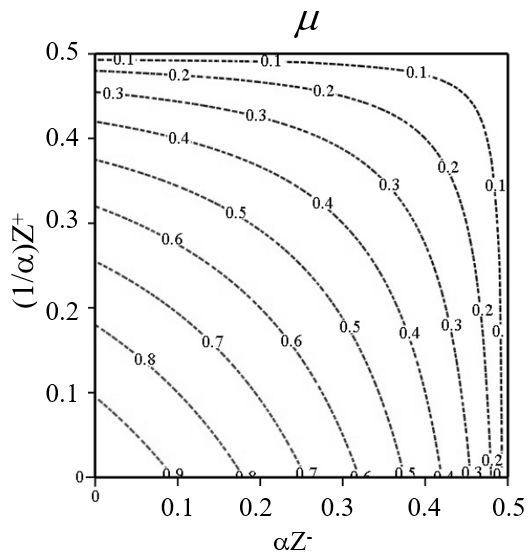}
\caption{\small Contour plot of the mass ratio Eq. \ref{masschange1} as a function of $z^+/\alpha$ and $\alpha z^-$.}
\label{fig:fig04}
\end{figure}

After determining the feasible region for the radiation emission energy, contour plots of the source mass relation, Eq. \ref{masschange1}, are as shown in Fig. (\ref{fig:fig04}). In this plot, the axes are expressed in terms of re-scaled values $z^+/\alpha$ and $\alpha z^-$ to provide a general view of the mass dependence on the emitted radiation. If, for example, $\beta=0$, the mass ratio $\mu\rightarrow 0$ as $z^{\pm}\rightarrow 1/2$ as the maximum value for the photon energy. 
\begin{figure}[!ht]
\centering
\includegraphics[width=0.4\columnwidth]{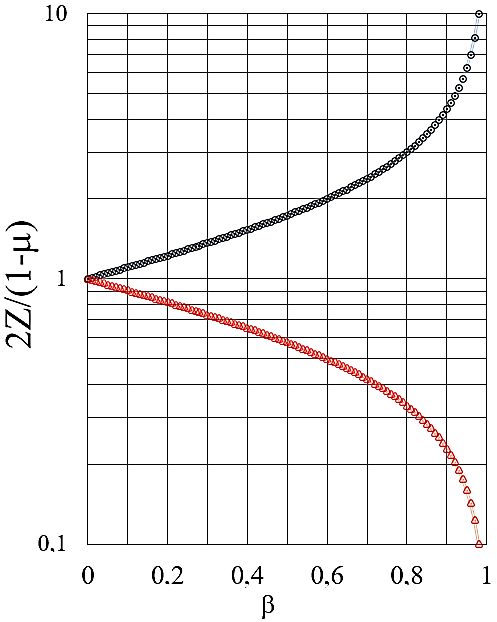}
\caption{\small Right (black) and left (red) radiated photon energies normalized by source mass lost $2z^{\pm}/(1-\mu)$ as a function of the velocity $\beta$ of a reference system.}
\label{fig:fig05}
\end{figure}

The radiated spectrum as a function of $\beta$, and the constrain $\beta=\beta'$ is shown in Fig. (\ref{fig:fig05}). The energies depicted in this plot correspond to the  classical Doppler shift ratios per unit of lost mass fraction of the source, $(1-\mu)/2$, as given by Eq. \ref{zs} for a cospectral emission in the source rest frame. In the low velocity limit, the photon energies are proportional to $\beta$, that is $z^{\pm}\approx 1/2(1-\mu)(1\pm \beta)$ while the high velocity limit with $\beta= 1-\epsilon$, $\epsilon\ll 1$, are $z^+\approx 1/2(1-\mu)\sqrt{2/\epsilon}$ and $z^-\approx 1/2(1-\mu)\sqrt{\epsilon/2}$.

\section{Practical example, graphical interpretation, and suggested problems}

In reality a source can emit a bunch of photons (or a beam) with arbitrary frequency distributions. The emission may involve unequal photon numbers and be called `anisotropic'. Anisotropic emissions may be responsible for unexplained behavior of spacecraft as observed in the anomalous acceleration in the 'Pioneer anomaly' \cite{pioneer}. Similarly, the emission may not be simultaneous in the source rest frame. A possible generalization of the dynamical Doppler shift with arbitrary intensities but still monochromatic beams is to assume energies $n^{\pm}h\nu^{\pm}$, with $n^{\pm}$ the number of emitted photons in each beam. These quantities are invariant upon a change of reference frame, or $n^{\pm}=n_0^{\pm}$. It is straightforward to show that for such a case, instead of Eqs. \ref{masschange1} and \ref{velgen}, the following relations should be used
\begin{equation}\label{masschange3}
\mu = \sqrt{\left(1-2n^+z^+/\alpha\right)\left(1-2n^-z^-\alpha\right)},
\end{equation}
and
\begin{equation}\label{velgen2}
   \beta' = \frac{\gamma\beta-(n^+z^{+}-n^-z^{-})}{\gamma-(n^+z^{+}+n^-z^{-})}.
\end{equation}
The new feasible mass domain is now dependent on the total number of photons in each beam, but essentially remains the same: $0\le n^+z^+\le \alpha/2$ and $0\le n^-z^-\le 1/2\alpha$ as suggested by Eqs. \ref{const1}.

It is instructive to apply the equations to a real system. Consider for example two 525 nm laser pens attached to each other. Each pen has $M=50$ g, and emits, for 7 days at the maximum power of 5 mW, two counter propagating laser beams. The system total mass is 100 g and the equivalent total energy released is 3.91$\times 10^3$ J. Each light beam contains $n=8.1\times 10^{21}$ photons carrying $3.73\times 10^{-19}$ J. The 7-days light beam stretches for 1212 A.U. (Astronomical Units) or 0.02 light-years from the source initial position. The dimensionless energies of each beam is therefore $z^{\pm}=6.73\times 10^{-13}$ and deplete the source mass by $\Delta\mu=1.34\times 10^{-10}$\%. Such small numbers make evident how large $Mc^2$ is in relation to typical emission powers of commercially available sources. In order to be effective, the radiation sources cannot be based on chemical processes, but on much more powerful ones - like nuclear reactors \cite{huth}. 
\begin{figure}[!ht]
\centering
\includegraphics[width=0.4\columnwidth]{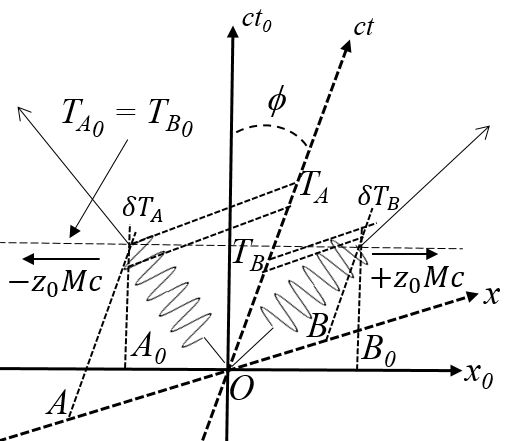}
\caption{\small Minkowski diagram for cospectral emission in the source rest frame, hence the energies are such that $z_0^+=z_0^-=z_0$. In this diagram $\tan\phi=\beta$, and  $\delta T_A$ and $\delta T_B$ are the projections of each beam's lengths onto the time axis of the moving frame.}
\label{fig:fig06}
\end{figure}

\subsection{Graphical interpretation}

The process of light emission by a moving source in 1-d may be illustrated on the Minkowski space-time diagram of Fig. (\ref{fig:fig06}) with a source resting in the $S_0$ system. In this frame, the two counter propagating photons with momenta $\pm z_0^{\pm}Mc$ (with $z_0^+=z_0^-$) will spread out on a light cone at 45$^\circ$ in relation to the orthogonal axis $x_0$ and $ct_0$. Sensors placed at  $A_0$ and $B_0$ on the $x_0$-axis with $\overline{OA_0}=\overline{OB_0}$ will detect the tip of the beam simultaneously (at time $T_{A_0}=T_{B_0}$). A moving frame is represented by a set of non-orthogonal axis sharing the same origin $O$. The source world-line will be represented by the segment $\overline{Oct}$ forming an angle $\phi$ with the  $ct_0$-axis so that $\tan\phi =\beta$. As it is clearly seen, the two space-time events will be first detected by a sensor located at $B$ and then at $A$. Both sensors are not equally spaced in relation to the point of emission. The beam heads will be detected at distinct times $T_A$ and $T_B$ on the $ct$-axis with $T_B < T_A$. The time interval between successive wave crests or troughs will be distorted, so that waves moving toward $+x$ will have higher frequencies than those going to $-x$. 
\begin{figure}[!h]
\centering
\includegraphics[width=0.4\columnwidth]{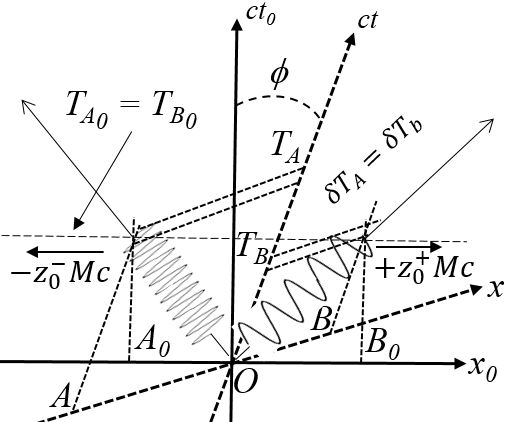}
\caption{\small Minkowski diagram for asymmetrical emission in the source rest frame, but, as observed from the right reference system,  $z^+=z^-$. Here $\tan\phi=\beta$. $\delta T_A$ and $\delta T_B$ are the projections of each beam's lengths onto the time axis of the moving frame.}
\label{fig:fig07}
\end{figure}

Given the time transformation between reference time frames, $S_0\rightarrow S$ , $t=\gamma[t_0+\beta x_0/c]$ \cite{french, dich}, so that the intervals calculated in each frame will be related by $\delta t =\gamma[\delta t_0+\beta \delta x_0/c]$. In Fig. \ref{fig:fig06}, $\delta t$ may be taken as $\delta T_A$ or $\delta T_B$ representing the projection of a given crest count on the moving frame time-axis. Dividing both sides by $N$ or the total number of crests counted by the sensors in each frame (which is invariant) we find $\delta t/N=\gamma[\delta t_0/N+\beta \delta x_0/(cN)]$. However, $\nu^{-1}=\delta t/N$ and $\nu_0^{-1}=\delta t_0/N$. Moreover, $\delta x_0/cN=(\delta t_0/N)[\delta x_0/(c\delta t_0)]=\delta t_0/N=\nu_0^{-1}$ since $\delta x_0=c\delta t_0$, the total length of $N$ crests during the interval $\delta t_0$. Therefore,  $\nu^+=\nu_0\alpha$ is the frequency measured for the right propagating photon by the sensor at point $B$. For the left propagating photon at point $A$ the same relations can be applied and we get $\nu^-=\nu_0/\alpha$. The relations Eqs. \ref{zs} for problem (a), Fig. \ref{fig:fig01}, are then graphically explained.

On the other hand, Fig. \ref{fig:fig07} represents problem 2(b), Fig. \ref{fig:fig02} when $z^+=z^-$. In the source rest frame, the two counter-propagating beams should be asymmetrically distributed in frequency so that their projections onto a particular moving frame time-axis becomes cospectral. As given by Eq. \ref{velgen}, this is only possible if the source velocity changes. Such a velocity change would be represented in Fig. \ref{fig:fig07} as a change in the inclination of the both time and space axis (to a new $\phi'$ with $\tan\phi'=\beta'$). The approach used to calculate the mass and velocity variations is completely general and carries an implicit assumption that the time interval of radiation emission is much shorter than any typical propagation times of the source as seen withing the observer time frame.   

\subsection{Suggested problems}\label{suggestedproblems}

In order to further strengthen the concepts, this section suggests six problems based on the presented discussion.

\begin{enumerate}
    \item Calculate the velocity of the reference frame $\hat{\beta}$ for which the source will be at rest after the emission of two photons with $z^{\pm}$ as
    \begin{equation}
        \hat{\beta}=\pm\frac{(z^+-z^-)}{\sqrt{1+(z^++z^-)^2}}.
    \end{equation}
    \item Show that, to first order in $z^{\pm}$, the mass ratio can be written in terms of $\alpha'$ as
    \begin{equation}
        \mu\approx \frac{1}{2}\left[ 3-\left(1+2z^-\alpha'\right)\left(1+\frac{2z^+}{\alpha'}\right)\right].
    \end{equation}
    \item Write the squared norm of the system 4-momentum, Eq. \ref{4mom}, after the photon emission in terms of the final source velocity, showing that it can be written compactly as
    \begin{equation}
        {\cal P'}\cdot{\cal P'}=-Mc^2\left(\mu+\frac{2z^+}{\alpha'}\right)\left(\mu+2z^-\alpha'\right).
    \end{equation}
    \item Show that, in Problem (3), ${\cal P'}\cdot{\cal P'}=-Mc^2$, or that mass is strictly conserved in the process. Discuss the meaning of this conservation in face of the reduction in the source mass $(1-\mu)M$.
    \item A moving source emits two cospectral counter propagating beams with $\nu^{\pm}=\nu$ and does not change its velocity after the emission. Show that this is only possible if the source velocity $\bar{\beta}$ is
    \begin{equation}
        \bar{\beta}=\frac{n^+-n^-}{n^++n^-},
    \end{equation}
    with $n^{\pm}$ the number of emitted photons in each beam.
    \item Show that, in the source reference system of Problem 5, the two beams have different frequencies given by
    \begin{equation}
    \begin{split}
         \nu_0^+ &=\nu\sqrt{\frac{n^-}{n^+}},\\
         \nu_0^- &=\nu\sqrt{\frac{n^+}{n^-}}.
    \end{split}
    \end{equation}
     In this case, no velocity change is observed in the source proper frame as well. Compare this situation with the one described in Section 2.3.
   % \item A source with velocity $\beta$ in a reference frame system emits two counterpropagating photons with $z^{\pm}=z$. Calculate $z^{\pm}_0$ in the source rest frame. Hint: Another equation is needed to find the two variables $z^+_0$ and $z^-_0$. 
\end{enumerate}

\section{Conclusion}

The Doppler shift is an intuitive phenomenon apprehended easily when an approaching siren is heard at distance. In ordinary optics, the Doppler shift is presented formally as relation involving the velocity of the radiation source and its proper frequency. Such relationship might give the impression that it is a purely kinematic expression as suggested by the sound equivalent. So, a question worth discussing with students is on the fate of the Doppler shift, because, according to Eqs. \ref{zs}, the emission frequencies depend on the final source mass and velocity which is true even in the cospectral case. In fact, by imposing $z^+=0$ and $z^-=0$ on Eqs. \ref{zs}, the only possible solution leads to $\mu=1$ or no mass change.

%The Doppler shift is a dynamical effect reducing the source mass which enters explicitly in the relations for the emitted radiation as shown in Eqs. \ref{zs}. 

 Using conservation of energy and momentum, which are central concepts in special relativity, this work emphasized the role of the mass-energy equivalence and mass conservation. But how is mass conserved? The suggested problem (4) clarifies the question. Problem (5) explores the dynamical case, when two counter-propagating beams are emitted with distinct frequencies in the source rest frame; however, they are detected as cospectral from a reference frame moving with velocity $\bar{\beta}$. 

Some interesting pedagogical consequences can be drawn from this study. For an isotropic source, the asymmetry in the forward and backward photon frequencies observed by a reference frame moving at velocity $v$ in relation to the source is not associated with any velocity change. However, it is possible to have a moving source emitting isotropic and cospectral radiation followed by an apparent velocity change which is very small, or of the order $z\beta$ as expressed in Eq. \ref{newvelinverteddoppler}. In the source reference frame however the emitted photons do not share the same frequency nor are emitted simultaneously as illustrated by the Minkowski diagrams of Fig. \ref{fig:fig06}. Notice however that the kinematic aspects of measurement process in distinct reference frames are bypassed by the  Lorentz invariance of Eqs. \ref{eqce} and \ref{eqcm}, which imply in relations among initial and final source velocities and photon energies only. This is an important pedagogical advantage of using conservation equations. 

The examples discussed here show the internal coherence of the relativity theory. In it, all concepts are interrelated: the necessary 4-vector invariance upon a change of reference system through the Lorentz transformation implies in the conservation of the new fundamental quantity, the 4-momentum. Mass is in fact conserved, but should be properly substituted or reinterpreted by the concept of energy which characterizes radiation while mass does not.

\section*{Acknowledgments}
The author would like to thank Christine F. Xavier for the help with the work. 

\section{Appendix}

First and second order derivatives of $\mu$  (mass ratio), Eq. \ref{masschange1}: 

\begin{equation}\label{eqdev1mu}
\begin{split}
    \frac{\partial \mu}{\partial z^+}&=-\frac{1}{\alpha}\sqrt{\frac{1-2z^-\alpha}{1-2z^+/\alpha}}\\ 
    \frac{\partial \mu}{\partial z^-}&=-\alpha\sqrt{\frac{1-2z^+/\alpha}{1-2z^-\alpha}},
\end{split}
\end{equation}

\begin{equation}\label{eqdev2mu}
\begin{split}
    \frac{\partial^2 \mu}{\partial z^{+2}}&=-\frac{1}{\alpha^2}\sqrt{\frac{1-2z^-\alpha}{(1-2z^+/\alpha)^3}},\\
     \frac{\partial^2 \mu}{\partial z^{-2}}&=-\alpha^2\sqrt{\frac{1-2z^+/\alpha}{(1-2z^-\alpha)^3}},\\
     \frac{\partial^2 \mu}{\partial z^+\partial z^-}&=\frac{1}{\sqrt{(1-2z^+/\alpha)(1-2z^-\alpha)}}.
\end{split}
\end{equation}

First and second order derivatives of $\beta'$ (final velocity), Eq. \ref{velgen}:

\begin{equation}\label{eqdev1beta}
\begin{split}
 \frac{\partial \beta'}{\partial z^+} &=\left(\frac{1}{\alpha}\right)
 \frac{(2z^-\alpha-1)}{[\gamma -(z^++z^-)]^2},\\
 \frac{\partial \beta'}{\partial z^-} &=\frac{\alpha(1-2z^+/\alpha)}{[\gamma-(z^++z^-)]^2},
\end{split}
\end{equation}

\begin{equation}\label{eqdev2beta}
\begin{split}
\frac{\partial^2\beta'}{\partial z^{+2}}&=\left(\frac{2}{\alpha}\right)\frac{(2z^-\alpha -1)}{[\gamma-(z^++z^-)]^3},\\
\frac{\partial^2\beta'}{\partial z^{-2}}&=\frac{2\alpha(1-2z^+/\alpha)}{[\gamma-(z^++z^-)]^3},\\
\frac{\partial^2\beta'}{\partial z^+\partial z^-}&=\frac{2[\beta\gamma-(z^+-z^-)]}{[\gamma-(z^++z^-)]^3}.
\end{split}
\end{equation}

\end{document}